\documentclass[journal=jctcce,manuscript=article,layout=twocolumn]{achemso}
\usepackage{amsmath}
\usepackage{graphicx}
\usepackage{dcolumn}
\usepackage{bm}

\def\beq{\begin{equation}}
\def\eeq{\end{equation}}
\def\bea{\begin{eqnarray}}
\def\eea{\end{eqnarray}}
\def\brcl{\begin{array}{rcl}}
\def\bccl{\begin{array}{ccl}}
\def\blcl{\begin{array}{lcl}}
\def\err{\end{array}}

\title{Molecular shape as a (useful) bias in chemistry}
\author{Guido Falk von Rudorff}
\email{guidofalk.vonrudorff@unibas.ch}
\affiliation{Institute of Physical Chemistry and National Center for Computational Design and Discovery of Novel Materials (MARVEL), Department of Chemistry, University of Basel, Klingelbergstrasse 80, CH-4056 Basel, Switzerland}

\begin{document}

\begin{abstract}
One of the molecular properties most intuitive to the human perception is the geometrical shape. However, when exploring a large chemical space the determination of shape needs to be automated. We present a fast and simple approach to identify a molecule as linear, planar, cube, cuboid, disk, elliptical disk, spheroid and sphere which is more fine grained than existing approaches. The method is applied to more than one billion molecules ranging from small organic molecules to whole proteins. The results show that current chemistry research is biased towards planar geometries. Moreover, we demonstrate that our molecular shape classification correlates with sought-after properties like the band gap, dipole moment, and heat capacity. This allows to increase the efficiency of molecular design studies by driving high-throughput-screening efforts towards desired values of molecular properties.
\end{abstract}

\maketitle

\section{Introduction}
The shape of molecules is an intuitive property which is of relevance e.g. in models of dipole interactions and polarization\cite{Onsager1936} or molecular packing and models of anisotropy\cite{Yokoyama2011,Mingos1991}. However, the subjective nature of an intuitive property calls for a classification which can be automated and scales to modern data sets with billions of molecules\cite{Ruddigkeit2012}. 

Previous approaches in assessing the distribution of shapes in chemical space include the calculation of the principal moments of inertia\cite{Gavezzotti1989,Browning2017}. While the principal moments of inertia can be evaluated in $\mathcal{O}(n)$, they are not able to distinguish between different three-dimensional shapes like cubes and spheres\cite{Sauer2003}. Since planar or linear molecules are over-represented in molecular databases\cite{Ruddigkeit2012}, three-dimensional molecules are interesting targets in compound design or screening efforts\cite{Lovering2009}.

Initially, the question of molecular shape was raised in the context of comparing two molecules. To compare molecules of different composition, models of the surface have been defined, either by Gaussian functions\cite{Haigh2005} or by means of a hard sphere being rolled around a molecule, the solvent accessible surface\cite{Connolly1983,Lee1971}. The van der Waals surface based on atomic radii is similarly compelling, but faces the challenge that the radii are not well-defined over a broad range of compounds\cite{Rowland1996}. All these shapes then could be used to align molecules\cite{Mansfield2002} or to compare volumetric overlap as a measure of similarity\cite{Haigh2005}. For a more complete overview, we point to a recent review\cite{Kumar2018}. However, these methods were neither designed for nor successful in classification of molecular shape in terms of simple geometric objects. More recently, the fast-growing field of machine learning models has developed a broad range of similarity measures that aim at the more general problem of comparing molecules not only on known properties like nuclear coordinates but also on inferred estimations of physical and chemical properties.\cite{vonLillienfeld2018}

The convex hull was considered as well for a similarity comparison between molecules. With success for clustering similar molecules\cite{Papadopoulos1991}, the need for numerically efficient approximations arose\cite{Varshney1994}. In order to follow the molecular shape more closely, some approximations tried to accommodate for special cases like non-convex molecules\cite{Chau1987}, all in the domain of pairwise molecular shape comparison. The convex hull also was employed to approximate the solvent accessible surface via triangulation\cite{Akkiraju1996, Sanner1996}. For molecular shape comparison, the concept of alpha shapes which encompass the convex hull as a special case, offers a adjustable method that can be efficiently implemented\cite{Edelsbrunner1983} but is strongly sensitive to slight variations in nuclear geometry\cite{Wilson2009}. From what can be gathered from literature, the convex hull is a computationally efficient tool, but does not capture enough of the chemical nature of molecular interaction to tell whether two molecules are similar to each other. In this work, we focus on the convex hull as a measure of similarity with simple geometric shapes which is a problem the convex hull is much more suited for, as will be shown.

\section{Methods}
The shape of a molecule is determined by nuclear coordinates alone. Therefore, a purely geometric measure of the overall shape is desirable. Since the internal structure of a molecule does not matter to the overall shape, the descriptor of the shape should depend on the surface of a molecule alone. This metric should be applicable to molecules of arbitrary size and computationally efficient. We suggest to use the convex hull of all nuclear coordinates, which can be computed in $\mathcal{O}(n\log h)$ with the number of atoms of a molecule $n$ and the number of atoms on the surface $h$. While calculating the convex hull is non-trivial, well-tested implementations are widely available which renders this approach easy to implement in practice.

\begin{figure}[!h]
    \centering
    \includegraphics[width=\columnwidth]{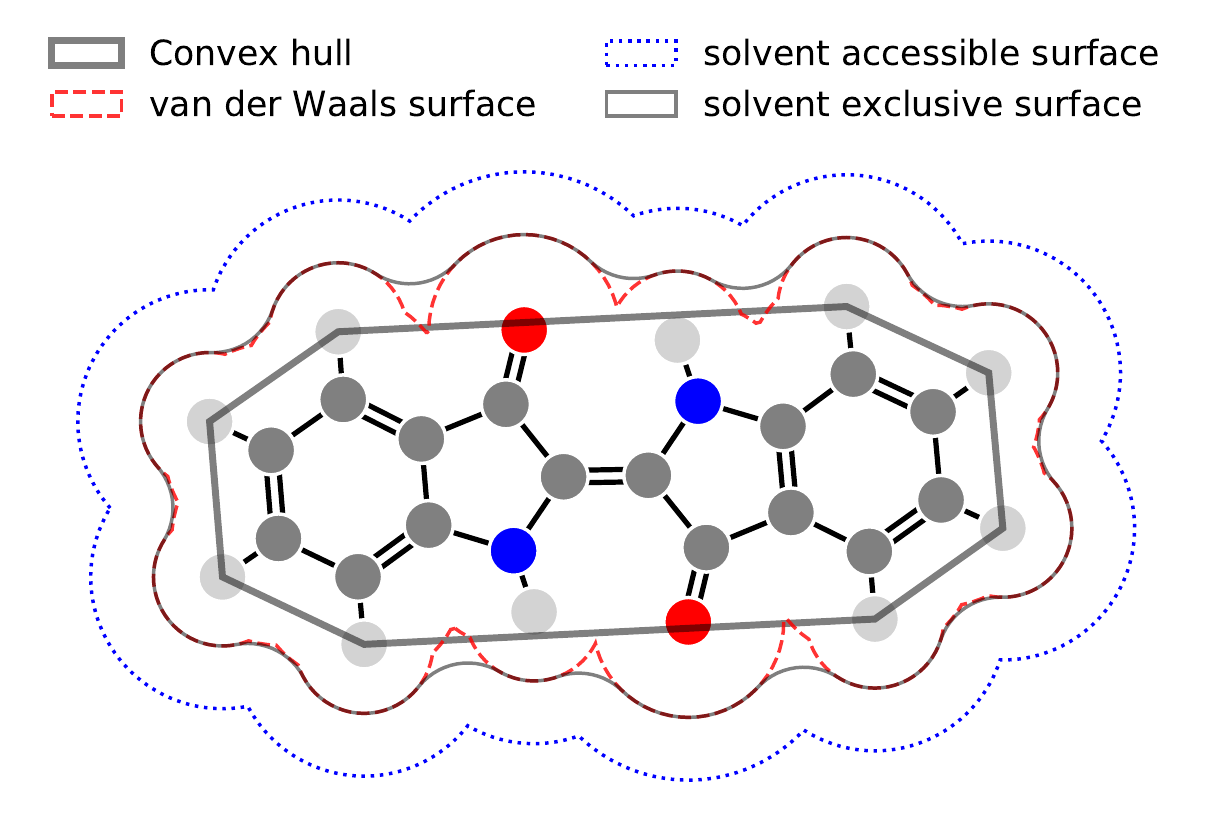}
    \caption{Convex hull for indigo compared to the van der Waals surface, the solvent accessible surface and the solvent exclusive surface.}
    \label{fig:convexhull}
\end{figure}

Figure~\ref{fig:convexhull} compares the convex hull in two dimensions to other molecular surface definitions. With the area and the volume of a convex hull, we can easily identify two cases: for a perfectly linear molecule, both area and volume of the convex hull are zero. For a perfectly planar molecule (like the one shown in Figure~\ref{fig:convexhull}), the volume is still zero, but the area is finite. This way, calculating the area and volume of the convex hull of a molecule is sufficient to tell apart whether a molecule is planar or linear.

This concept can be pushed further by seeing that the convex hull satisfies particular area to volume relations for certain geometric shapes. Exploiting this allows to rapidly classify a molecular configuration into similar overall geometric shapes. To that end, we propose to use the scaled specific surface area ($S$) 
\begin{align}
    S \equiv \frac{A}{V^{2/3}}
\end{align}
where $A$ and $V$ are the area and volume of the convex hull of all nuclear coordinates, respectively. This way, $S$ is unitless and independent of the size of the molecule being tested. Table~\ref{tab:sssa} shows the analytic values for various relevant geometric shapes. From these equations, it becomes clear that the area of the tested geometric shapes can be expressed as some function of $V^{2/3}$, hence $S$ is a convenient scalar measure.

\begin{table*}[t!]
    \centering
    \begin{tabular}{l|cccc}
         Shape & Volume $V$ & Area $A$ & Typical Range & S  \\\hline
         linear & 0 & 0 & -- & $\infty$ \\
         planar & 0 & finite & -- & $\infty$ ($>$28) \\
         elliptic disk (1:x:y) & $4\pi xya^3/3$ & $\simeq 4\pi\sqrt[8/5]{(x^{8/5}+y^{8/5}+(xy)^{8/5})/3}$ & $x \in [1.5, 8], y=0.2$ &11.4-19.4\\
         circular disk (1:x) & $\pi a^3x$&$2\pi a^2(1+x)$ & $x\in[3, 5]$ &8.1-10.3\\
         cuboid (1:x:y) & $xya^3$ & $2(y+x+xy)a^2$ & $x=1, y\in[2, 5]$ &6.3-7.5\\
         cube & $a^3$ & $6a^2$ & -- &6\\
         spheroid (1:x) & $4\pi a^3x/3$& $2\pi a^2 \left(1+\frac{x^3}{x^2-1}\arcsin(1-x^{-2})\right)$ & $x\in[1.5, 3]$ &4.9-5.7\\
         sphere& $4\pi a^3/3$ & $4\pi a^2$ & -- &4.8\\
    \end{tabular}
    \caption{Scaled specific surface area for various geometric shapes. In all cases, the characteristic length unit is $a$. For shapes with multiple characteristic lengths, the ratio of which is specified via $x$ and $y$. Linear and planar molecules can be distinguished by their area. Numeric values for $S$ given are derived from inserting the typical parameter range $x, y$ into the expressions for volume and area. In our implementation, the closest classification is chosen. Boundaries in Figure~\ref{fig:combined} are drawn using these value of $S$.}
    \label{tab:sssa}
\end{table*}

Using the qhull code\cite{Barber1996} to obtain the convex hull and OpenBabel 2.4.0\cite{OBoyle2011} to obtain geometries from SMILES where necessary or RDKit to parse data files, we applied the method to GDB13\cite{Blum2009}, the PubChem3D database \cite{Kim2018,Bolton2011}, and the Protein Data Base (PDB)\cite{Berman2000}. GDB13 contains 939 million molecules with up to 13 heavy atoms. Since these molecules are generated systematically, they are less affected by selection bias and therefore serve as a measure of how molecular shapes are distributed for small organic molecules. PubChem3D is more comparable to GDB13 in terms of molecule size and reflects the compound space of small molecules that have been research subject in theoretical or experimental work. It is more interesting to compare molecular shapes for entries of the PubChem3D database of known molecules against those from the GDB13 database of possible molecules, since this allows to unravel any bias introduced by the focus of research studies.

The link between overall molecular shape and chemical properties is investigated by using geometries and properties calculated at B3LYP level of theory as part of the QM9 database\cite{Ramakrishnan2014} which again covers a subset of possible small organic molecules. Since this database (unlike GDB13) is not designed to be exhaustive for some chemical compound space, only in-sample comparisons of property distribution can be performed.

As a benchmark of the performance of the method, the Protein Data Base has been included where individual molecules typically consist of several thousand atoms. If an entry in PDB contains multiple structures of the same protein, all of them have been treated as a single entry. Only the core protein structure has been included in the shape assessment: solvent molecules have been stripped from the geometry. While the atom count of the molecules intuitively suggests that spheroid configurations should dominate PDB entries, this database demonstrates the computational efficiency of the method and validates the shape classification in the limit of large molecules.

\section{Results}

\begin{figure*}
    \centering
    \includegraphics[width=\textwidth]{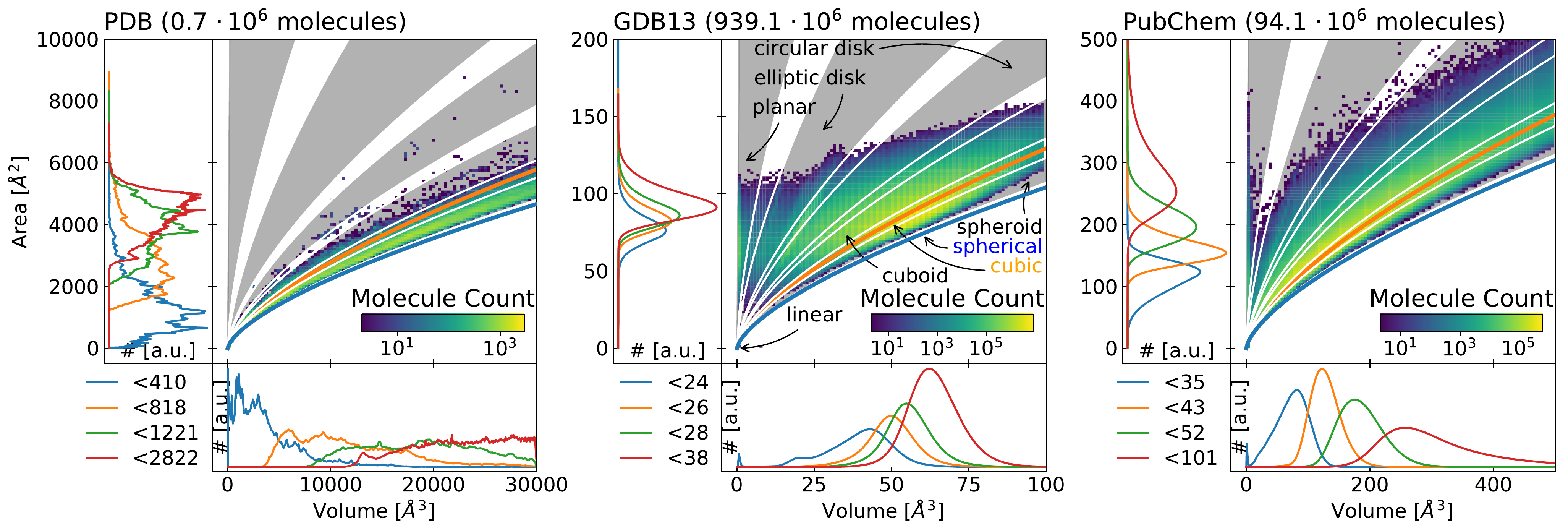}
    \caption{Distribution of convex hull volume and convex hull area as a measure of molecular shape for one billion molecules in three databases: PDB, GDB13, and PubChem3D. White areas in heatmaps correspond to combinations of surface area and volume that are not found for a single molecule in the respective database. Shaded areas in heatmaps denote the regions of the different geometric shapes, as specified in Table~\ref{tab:sssa}. Histograms below of and left of heatmaps show the distribution of volume and area for the molecules in the four atom number quartiles.}
    \label{fig:combined}
\end{figure*}

Figure~\ref{fig:combined} shows the heatmap of molecular volume and molecular area as obtained from the all-atom convex hull for the various databases. As expected, the spherical shape is the lower bound in this diagram, since it has the lowest area for a given volume of all possible geometric shapes. Planar and linear molecules are only found in databases which contain small molecules, since these configurations are unlikely to be stable with increasing atom count. This is highlighted by the volume histogram for GDB13 where only the subset of smallest molecules shows a peak close to zero volume, i.e. for linear or planar molecules.

While the distinct classes of molecular shapes are present for both PubChem3D and GDB13, it is remarkable that PubChem3D has a substantial bias towards planar molecules as compared to the distribution found from enumeration of possible molecules in GDB13, which might be due to 2D representations being more intuitively accessible. The bias manifests itself in higher number of compounds of large area for a given volume, in particular in the range of 30-40 atoms. Since GDB13 enumerates molecules up to 13 heavy atoms, the compounds of PubChem3D need to have a higher number of heavy atoms but be of comparable volume but larger area. This points towards compounds that are planar or disk-shaped but have less saturating hydrogen atoms than those from GDB13. 

In general, both GDB13 and PubChem3D feature all molecular shapes that can be separated with our approach. From both datasets, it becomes clear that spherical configurations are hardly reached for larger molecules. This is plausible since only few cases like fullerenes conform to this shape.

For large molecules as those in PDB, the dominating molecular shape is a spheroid. This is a plausible result of folding since the weak attractive dispersive and electrostatic forces between molecular chains stabilize a compact geometry. This trend is already visible for smaller molecules, e.g. in GDB13 as the surface area increases only very slowly with atom count whereas the volume increases faster which overall shifts the shape distribution towards more spherical configurations.

\begin{figure*}
    \centering
    \includegraphics[width=\textwidth]{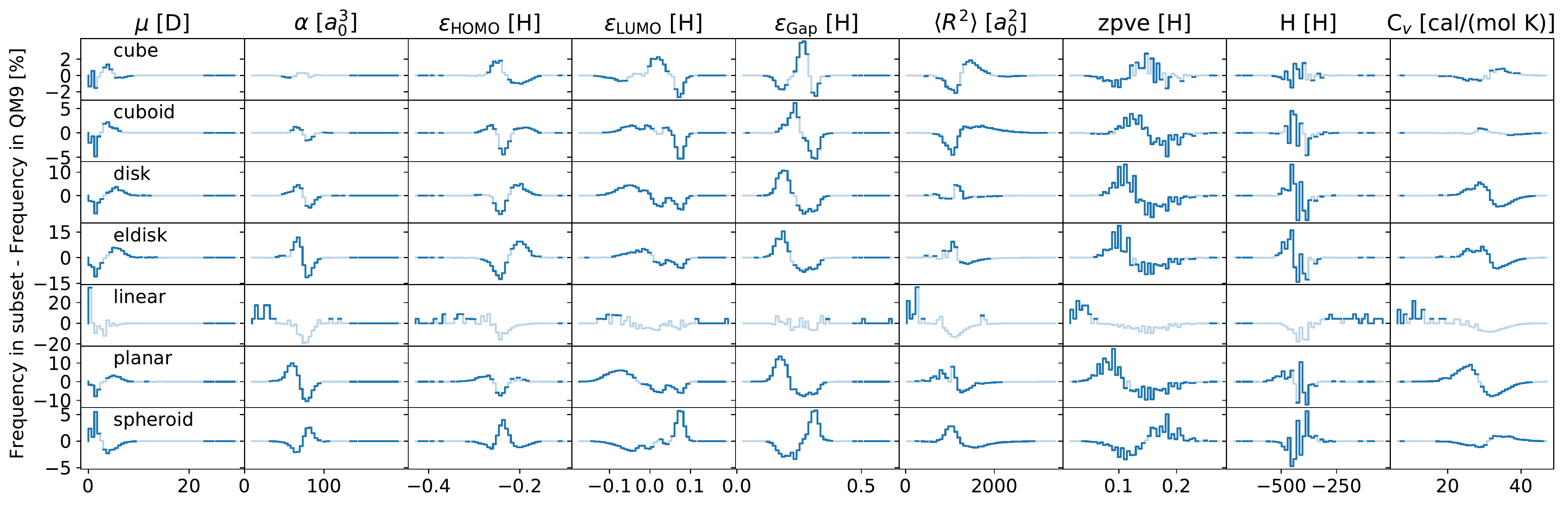}
    \caption{Distribution of properties (columns) for molecules of different shapes (rows) from the QM9\cite{Ramakrishnan2014} database compared to the overall distribution in the database. Dark lines are statistically significant, light lines could be explained by sample bias. Shown are dipole moment $\mu$, isotropic polarizability $\alpha$, energy of HOMO $\epsilon_\mathrm{HOMO}$, energy of LUMO $\epsilon_\mathrm{LUMO}$, band gap $\epsilon_\mathrm{Gap}$, electronic spatial extent $\langle R^2\rangle$, zero point vibrational energy zpve, enthalpy H, and the heat capacity C$_v$. The internal energy at 0\,K U$_0$, the internal energy at 298.15\,K U, and the free energy at 298.15\,K G are virtually identical to the histogram of the enthalpy H.}
    \label{fig:propertygrid}
\end{figure*}

Having a fast and rigorous method of classifying molecular shapes as presented in this work can help screening efforts attempting to link structural features and physical properties. Using the QM9 dataset\cite{Ramakrishnan2014}, we demonstrate how this could be done. To this end, we calculate the convex hull of all molecules in QM9 and classify their shape as mentioned above. For each physical property of a molecule, e.g. the dipole moment, we therefore obtain a distribution of values for the whole QM9 dataset and for the subsets of identical shape classification. If there is no correlation between shape and molecular property, the distribution of, say, dipole moments in the whole dataset and in all subsets should be identical. Figure~\ref{fig:propertygrid} shows the resulting differences between the full dataset and the shape-grouped subsets. 

It is important to note, however, that this procedure is susceptible to a statistical effect, the sample bias. When drawing a subset of molecules that conform to a certain geometric shape from a larger database, the different property distribution of the subset is subject to a sampling error. In order to tell apart when the differences are large enough to be likely correlated with molecular shape, we use the following sampling error estimation. 

For any $i$-th histogram interval $[x_i, x_{i+1})$, the full dataset of length $N$ has a certain count $C_i$ of entries within this interval. Therefore, randomly drawing a subset of size $n$ has a probability $p=C_i/N$ to pick one entry within this interval. The subsequent random selection is a Bernoulli process, resulting in a Binomial probability distribution $\mathcal{B}$ of the outcomes $c_i$
\begin{align}
    \mathcal{B}(c_i) = \binom{n}{c_i}p^{c_i}(1-p)^{n-c_i}
\end{align}
The standard deviation $\sigma$, i.e. the expected sampling error, is given as
\begin{align}
    \sigma \equiv \sqrt{np(1-p)} = \sqrt{\frac{nC_i}{N}(1-C_i/N)}
\end{align}
In the histograms in Figure~\ref{fig:propertygrid}, we highlight any bin where $|c_i - nC_i/N| > 3\sigma$.

In Figure~\ref{fig:propertygrid}, we show that many properties of interest correlate with molecular shape. Dipoles of spheroidal molecules are shifted towards lower values while this is not the case for disk-like shapes where the inverse effect is even stronger. This suggests that the distribution of partial charges such that they give rise to a strong dipole moment is more common and easier to achieve for disk-like molecules. In the context of QM9, this points towards five-rings or six-rings with branching chains. 

For polarizability, one would expect that more compact configurations like cuboid shapes coincide with low polarizability. While disk-like and planar configurations should have directions of stronger polarizability, the isotropic polarizability as available in QM9 should not reflect this sense of directionality. And indeed, Figure~\ref{fig:propertygrid} shows a shift towards higher polarizabilities only for spheroidal configurations. 

In the context of organic semiconductors, it is not only the band gap that is of interest but also the relative band alignment, i.e. HOMO and LUMO positions\cite{Helander2008}. With organic semiconductors being commonly planar in practice\cite{Yokoyama2011}, it is particularly relevant to see whether molecular shape could be used to adjust band alignment. From Figure~\ref{fig:propertygrid}, one can see that the variance of HOMO positions is reduced for spheroidal molecules. If a particular application demands for higher-lying HOMO levels, disk-like molecules have a higher probability of exhibiting this behaviour, while cubic or cuboid molecules tend to have lower lying HOMO levels. For the LUMO, spheroidal molecules exhibit the strongest push towards higher values, while all other shapes seem to tend to lower lying LUMO energies. It is worth noting however, that Figure~\ref{fig:propertygrid} does not consider conditional probabilities: having one molecular shape with typically higher HOMO and lower LUMO does not necessarily mean that all its molecules have a smaller band gap. This can be seen from the band gap distribution where the differences in molecular shape are significantly stronger pronounced, e.g. in the disk-shape case where a broad difference in LUMO distribution gives rise to a sharp and strong shift towards lower band gaps. This knowledge can help in molecular design problems relevant to absorption, i.e. in layered photovoltaic materials.

With the electronic spatial extent as a measure of molecular volume, four shapes are clearly different from the overall QM9 dataset. Spheroidal molecules have much less variance in the extent then the overall dataset, which is reasonable as sphere-like objects for a set number of heavy atoms should span similar volumes with their electronic density. On the contrary, cubic or cuboid molecules require a larger volume than the rest since their shape is less compact than the one of a sphere. Planar objects have a lower spatial extent, since they leave out one dimensions compared to other molecules. With the electronic spatial extent being a geometric measure, the results for this property are not surprising, but are more of a validation and plausibility argument of the proposed method.

For completeness, Figure~\ref{fig:propertygrid} shows the zero-point vibrational energy and internal energies for the QM9 dataset, even though these quantities are rarely target of an optimization in materials design, they yield plausible trends, for example a lower ZPVE for linear molecules and an increased one for spheroidal molecules. From the thermodynamic quantities in the dataset, the heat capacity is still quite interesting in the design context: disk-like and planar compounds exhibit a lower-than-typical heat capacity, while more isotropic shapes like cubes or spheroidal molecules have a higher probability for larger heat capacities.

\begin{figure}
    \centering
    \includegraphics[width=\columnwidth]{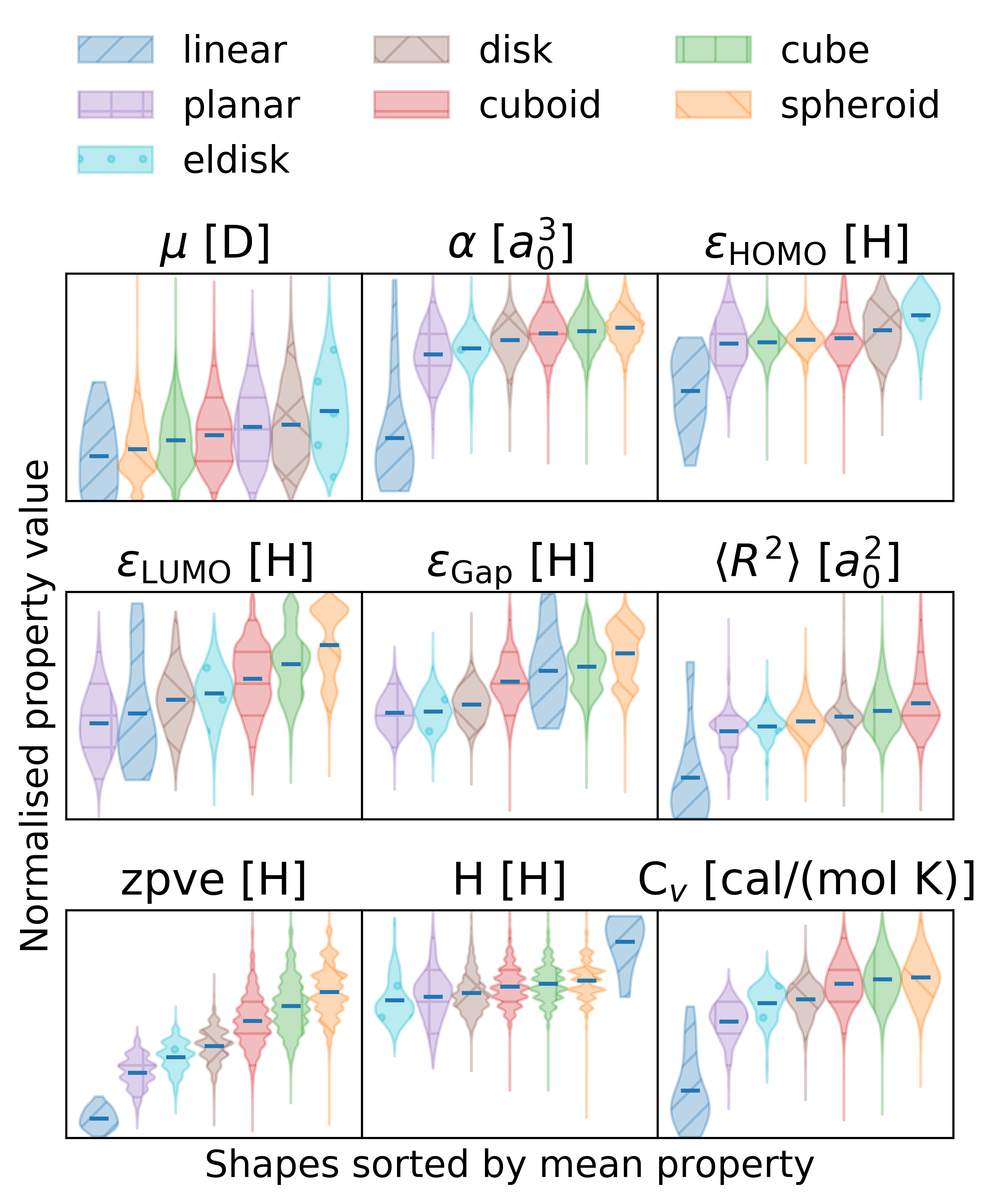}
    \caption{Distribution of QM9 properties by molecular shape obtained by kernel density estimation. For each property, shapes are sorted by the corresponding mean property value (short horizontal bars). All distribution plots normalized to ease comparison.}
    \label{fig:ordering}
\end{figure}

One must never forget to look at the aim of a matter: all these examples suggest that molecular shape could be used as one criterion in materials design efforts, where it is key to focus the finite computational resources on molecules that have a higher probability of having the desired target properties. Since the number of potential molecules is so gigantic, a slight but notable shift in success probabilities improves high-throughput-screening efficiency without side-effects. 

To allow a direct comparison of the bias in property value based on the molecular shape, Figure~\ref{fig:ordering} shows the distribution of each property for each shape one one common scale. For all properties under consideration, molecular shapes have a strong impact on the distribution. Most prominently this is the case for the heat capacity and for the band gap, both relevant properties in the materials design community. While the individual properties have been discussed in detail above, Figure~\ref{fig:ordering} highlights the fact that molecular shape presents a bias for a wide range of physical properties, hence, is of general value.

It is worth noting though that the QM9 database only contains one conformer per molecule. While very small (less than five atoms) and very large molecules (on the PDB scale) have low-energy conformers of very similar shape, intermediate molecular sizes might give conformers of vastly different shape, so it would be interesting to assess whether the trend described in this work holds for conformers of the same molecule as well. Based on the numerical evidence of this work and with QM9 containing any conformer, not necessarily the minimum energy one, it is likely that this trend would be confirmed. A reliable estimate, however, requires a nearly exhaustive dataset of conformers for some molecules. An effort to this end is currently in preparation.

As an additional dimension, the atom number density of the convex hull could be taken into consideration. A major difference between surface estimation methods like the solvent accessible surface area and the approach proposed in this work is the treatment of cavities. By definition, the convex hull does include cavities in the total volume. This means that molecules with a higher number of cavities or larger cavities feature a lower atom number density. For larger molecules like proteins, the atom number density is more a descriptor of surface roughness, while for small molecules this is more of a way to identify cavities. This is simply because small molecules do not have enough atoms to form a rough surface, but they do have enough atoms to form cavities, for example like fullerenes. Consequently, the atom number density of small molecules could be used as screening parameter to identify small molecules with a cavity or proteins featuring particularly many (or few) pockets or channels, similar to approaches where the convex hull has been used starting point for pocket searches\cite{Meiera, Petrek2007,Connolly1991}.

While the proposed method is intuitive and easy to implement, there are edge cases where the shape classification is not as accurate. Most notably in small molecules, where most of the atoms reside on the surface of the convex hull, small changes in nuclear coordinates can give rise to a strong response in both surface area and volume. This is typically the case for molecules of four atoms in total, as molecules of three atoms are always planar. In this case, the cases linear, planar and 3D can be distinguished, but not more.

It is worth noting that the rescaling trick in the definition of $S$ works for most but not for all geometrical shapes. The surface area of a cone cannot be expressed as a prefactor to $V^{2/3}$. This is of limited relevance for shape classification though, since the thin tip of the cone is implausible for a molecular structure.

In an actual implementation, perfectly linear and planar molecules are edge cases since the convex hull becomes degenerate. We suggest to circumvent this by random displacements of 0.001\,\AA\, which does not change the overall result, but satisfies most convex hull solvers.

\section{Conclusion}
We have shown that the area and volume of the molecular convex hull can be used to classify molecular shapes. With this descriptor, we can assign one of the following eight molecular shapes: cube, cuboid, disk, elliptic disk, linear, planar, spheroid, and sphere. The scalar metric is simple, well defined and easy to implement.

Using this classification, we have shown that some relevant molecular properties like dipole moment, polarizability, HOMO, LUMO, band gap, and heat capacity correlate with the molecular shape. These correlations can be used to refine high-throughput materials design studies and increase the number of relevant molecules they find at the same computational cost.

The main limitation is that for very small molecules (i.e. $4-5$ atoms) the molecular volume is too heavily dependent on the actual nuclear configuration such that we can only classify linear, planar, and 3D structure.

The descriptor suggested and validated in this work is suited as a descriptor of the field of cheminformatics or as an input to machine learning models for rational compound design and allows to quickly link geometrical shape and molecular properties. All code this work is based on is freely available online\cite{github}.

\begin{acknowledgement}
This work was supported by a grant from the Swiss National Supercomputing Centre (CSCS) under project ID s848. 
Some calculations were performed at sciCORE (http://scicore.unibas.ch/) scientific computing core facility at University of Basel.
We acknowledge support by the Swiss National Science foundation (No.~PP00P2\_138932, 407540\_167186 NFP 75 Big Data, 200021\_175747, NCCR MARVEL).
GvR would like to thank Prof. Anatole von Lilienfeld for support and helpful discussions as well as Anders S. Christensen and Jimmy Kromann for comments on the manuscript.

\end{acknowledgement}

\bibliography{main,online}

\end{document}